\documentclass[11pt]{article}

\usepackage{graphicx}
\begin{document}

{\scriptsize \em
20th.October 2004}

\noindent
Philosophical Magazine, to be published.
\begin{center}
{\large{\bf Analysis of room-temperature results on normally 
conducting and superconducting channels through polymer films}}
\vskip 2em
{By D.M. EAGLES$^{1}$}

\vskip 1em
{19 Holt Road, Harold Hill, Romford, Essex RM3 8PN, England}
\vskip 2em

\vskip 2em
\noindent
{ABSTRACT}

\end{center}
\begin{quote}
There are strong reasons from dc and pulsed-current measurements, and
from thermal conductivity results, for thinking that narrow channels
through films of oxidised atatic polypropylene (OAPP) are
superconducting at room temperature.  It is thought that the conducting
channels, with diameters less than or of the order of a micrometre, are
composed of smaller nanofilaments, with diameters of the order of a
nanometre.  In the present paper a possible explanation is given of
measurements which show that the average resistance of
non-superconducting channels through films increases with film
thickness more slowly than linearly.  This result is interpreted in
terms of how the Bose condensation temperatures of bosons in arrays of
nanofilaments depend on the length and numbers of filaments, and
examples are given of parameters of the arrays which could explain the
data.  The dispersion for the bosons is assumed to consist of a sum of
linear and quadratic terms, which is an approximate type of dispersion
reported for Cooper pairs.  In order to fit the data with the model
used, it is necessary to suppose that values of superconducting $T_c$
for channels composed of large numbers of filaments are only slightly
above room temperature.  It is argued that larger $T_c$'s reported in
1989 when currents of 0.5 A are passed through channels may arise
because (a) currents concentrate in a subchannel of smaller width than
the original channel, and (b) current-current interactions draw the
filaments of the subchannel sufficiently close together to increase the
transverse bandwidth and $T_c$ in the model by the required amount.

\end{quote}


\noindent
Running title:  Normally conducting and superconducting channels
through polymer films.

\noindent
$^{1}$   e-mail: d.eagles@ic.ac.uk

\begin{center}
{\S 1. INTRODUCTION}
\end{center}

The first indication that narrow channels through films of oxidised
atactic polypropylene (OAPP) might be superconducting at room
temperature came from a finding from resistance measurements with
microprobes a few $\mu$m in diameter at various points on the top
surface of films of OAPP on a conducting substrate.  Some points showed
very low resistance independent of the film thickness and about equal
to that of a contact put directly on the substrate (see figure 1, taken
from figure 17 of Grigorov and Smirnova 1988).  (See also Enikolopyan {\em et
al.} 1989, Arkhangorodski\u{i} {\em et al.} 1989).

More convincing evidence came in 1990 and 1991 from three types of
results:  \newline (i) Estimates of lower limits on conductivity for
channels several orders of magnitude greater than that of copper by
direct (Arkhangorodski\u{i} {\em et al.} 1990) and indirect (Demicheva
{\em et al.} 1990) methods; \newline (ii) Destruction of
superconductivity by non thermal means by pulsed currents, with
critical current densities of the order of $5\times 10^9$ A cm$^{-2}$
(Demicheva {\em et al.} 1990); and \newline (iii) The finding of
negligible electronic thermal conductivity in the channels, violating
the Wiedemann-Franz law by several orders of magnitude (Grigorov {\em et
al.} 1991).

Some unusual magnetic effects (Enikolopyan {\em et al.} 1989, Grigorov
{\em et al.} 1993, Rogachev and Grigorov 2000) involving large diamagnetism
have been seen, and are probably associated with channels which form
closed loops.  A possible interpretation of large diamagnetism in low
magnetic fields at room temperature in some samples has been given
recently by the present author (Eagles 2002).  Other magnetic effects
such as superparamagnetism and metamagnetism (Smirnova {\em et al.}
1988, Enikolopyan {\em et al.} 1989, Grigorov {\em et al.} 1996, Rogachev
and Grigorov 2000) are thought to be associated with conducting channels
which do not form closed loops.  

The channels have maximum diameters of the order of 1 $\mu$m (Demicheva
{\em et al.} 1990), but for theoretical reasons (Grigorov 1990, 1991,
Grigorov {\em et al.} 1990) it is thought that the channels are
composed of large numbers of smaller nanofilaments, known as superpolarons,
of diameters of the order of one to a few nm.  The only two types of
published theory for superconductivity in the channels of which the
author is aware (Eagles 1994a, 1994b, Grigorov 1998) make use of the
assumption that the channels are composed of many quasi one-dimensional
filaments.  The present author (Eagles 1998a) and Grigorov and
coworkers (see e.g. Grigorov and Rogachev 2000) have very different
suggestions as to the origin of the magnetism in the non-closed-loop
channels, although both depend on the above assumption.

\begin{center}
{\S 2. SUPERCONDUCTING AND NORMALLY CONDUCTING CHANNELS} 
\end{center}

When the resistance is measured between a conducting substrate and
microprobes of diameters of approximately  10 $\mu$m placed on top of OAPP
films, distribution functions of the resistance for two different
thicknesses of film and for contacts put directly on the substrate are
shown in figure 1, taken from Grigorov and Smirnova (1988).
\begin{figure}
\rotatebox{270}
{\centerline{\includegraphics[width=12cm]{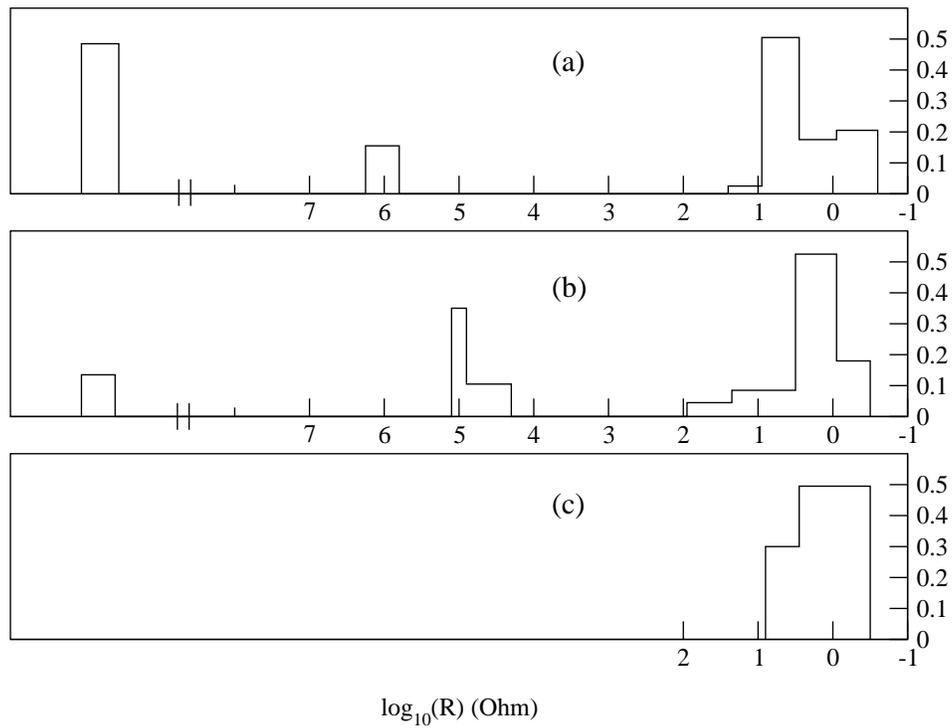}}}
\caption
{Distribution function for resistance $R$ at room temperature
from microprobes to a metallic substrate: (a) For a sample of oxidised
atactic polypropylene (OAPP) of thickness 50 $\mu$m, (b) for OAPP with
thickness 0.3 $\mu$m, and (c) for the metallic substrate without any
polymer film. [Reproduced approximately from figure 17 of Grigorov and
Smirnova (1988)].}
\end{figure}

Note that for the films there are three types of points (i) insulating,
(ii) mediumly conducting, and (iii) highly conducting.  For the mediumly
conducting points, the lowest resistance for the 50 $\mu$m film is
slightly more than 30 times as large as the lowest resistance for the
0.3 $\mu$m film, implying that the cross section of channels associated
with these points is about five times larger than for the 50 $\mu$m film
if the resistivity is the same for the material of these channels in
both films.  In a later paper (Enikolopyan {\em et al.} 1989), the
departure from linearity is smaller, and, for the results presented
there, we can deduce that the cross section of channels with the lowest
normal resistance in the thinner films is only about twice as large
as  the corresponding cross section for the thinner films.  Later in
the paper we use parameters which give a value of four for this ratio,
in between the 1988 and 1989 results.

The channels associated with the highly conducting points are probably
superconducting (see the Introduction), and in this paper we explore
the hypothesis that the reason for the apparently larger cross section
of the  normally conducting channels with the lowest normal resistance
in the thicker films is that a larger cross section is required for
superconductivity to occur at a given temperature in channels of larger
lengths.  The specific model which we use to examine this hypothesis
is a model of charged bosons on an array of quasi-one dimensional
filaments, with a boson dispersion consisting of a combination of
linear and quadratic terms, which is found to be a fair approximation
for the dispersion of Cooper pairs (Adhikari et al 2000).  Although a
Cooper-pair model gives unrealistically small binding energies for
large Fermi energies or for weak coupling because of neglect of what
happens below the Fermi surface, we think it may be more realistic
when pair binding energies are large compared with the Fermi energy
as in that case what happens below the Fermi energy is less
important.  Even for the case of large Fermi energies, although the
Cooper-pair binding energies are too small, it is possible that the
Cooper-pair model may give the correct form of the wave-vector dependence
of pair binding energies relative to those for zero wave vector.

Our calculated values of $T_B$ will be close to superconducting
transition temperatures $T_c$ if and only if the pairs do not break up
at a temperature lower than $T_B$, i.e. if we are on the Bose-gas
side of the BCS-Bose gas transition.  This transition was first studied
by Eagles (1969).  In the BCS theory the pair binding energy is
$3.5k_BT_c$, and is somewhat larger in its strong-coupling extensions.
Thus a necessary condition for our theory to be valid is that the
binding energy of pairs should be larger than 3.5$k_BT_B$,
where $T_B$ is calculated neglecting pair break-up.

We also mention that, for arrays with small numbers of nanofilaments,
the resistance at the condensation temperature cannot be expected to be
zero, because of either thermal (Langer and Ambegoakar 1967, McCumber
and Halperin 1970) or quantum (Zaikin {\em et al.} 1997, Zhao 2003) phase
slips of the order parameter.  The ultra-high conductivity of several
orders of magnitude larger than that of copper at room-temperature
mentioned in the Introduction has only been established for relatively
large arrays of nanofilaments forming channels of diameter of the order
of a micrometer, for which effects of phase slips are expected to be
negligible.  For smaller arrays the only experimental lower limits on
the conductivity are much lower and only approximate, based on the
lack of noticeable length dependence of the resistance of the channel
plus contact resistance of the highly conducting channels as indicated
in figure 1.

\begin{center}
{\S 3. CONDENSATION OF BOSONS IN AN ARRAY OF NANOFILAMENTS}
\end{center}

In this section we use an idealised model of conducting channels
composed of non-interacting bosons on a square array of quasi
one-dimensional filaments  of length $L$ and transverse dimension $D$
measured from point half a filament separation $a_T$ beyond the centers
of the edge filament rows.  The filaments are assumed to be
sufficiently narrow that the bosons in a filament can be assumed to be
in the lowest possible state with respect to motion perpendicular to
the filaments at all temperatures of interest.    We suppose that boson
wave functions vanish at the edges of the array, and neglect any
periodicity in the potential along the length of the array.  Then we
suppose that the boson states in the array can be characterised by  the
magnitudes of wave numbers $K_1$, $K_2$ and $K_3$ in the two transverse
and longitudinal directions, where $K_1$ and $K_2$ take on discrete
values ranging from $\pi/D$ to $n_T(\pi/D)$, and $K_3$ takes on
values of $k\pi/L$, where k runs from 1 to $\infty$.     
For computational purposes we approximate the sum to infinity by a finite
sum with a suitably high upper limit. 

Based on the fact that the dispersion relation for Cooper pairs can be
approximated as a combination of linear and quadratic terms
(Adhikari {\em et al.} 2000) we assume that the boson energies
$E_B(K_1,K_2,K_3)$ measured from the lowest state can be written as
\begin{eqnarray} 
E_B(K_1,K_2,K_3) =&
A_L[(\pi/L)^2(k^2-1)+(d_L/a_T)(\pi/L)(k-1)]\nonumber\\
&+[W_T/(1+s_T)][(i^2+j^2-2)/n_T^2+s_T(i+j-2)/n_T].
\end{eqnarray} 
Here 
\begin{equation}
A_L=\hbar^2/2M_L,
\end{equation}
where $M_L$ is the boson mass in the longitudinal direction, $a_T$ is
the interfilament separation, $d_L$ is a dimensionless constant
characterising the ratio of the linear to the quadratic terms in the
longitudinal boson dispersion relation, $s_T$ is the ratio of the
contributions of the linear and quadratic terms in the dispersion to
the transverse bandwidth $W_T$ (in a direction parallel to one of the sides of the cross section of the channel), and
\begin{equation}
p=M_L/M_T,
\end{equation}
where $M_T$ is the boson mass in the transverse direction.  $W_T$, $s_T$,
$A_L$, $p$ and $a_T$ are related by
\begin{equation}
W_T/(1+s_T)=p A_L(\pi/a_T)^2.
\end{equation}
We measure $W_T$ from hypothetical levels with $i=j=0$ in order to
avoid having a dependence of $W_T$  on $n_T$.  We will discuss values
of the parameters $d_L$ and $s_T$ in the next section.

The case of zero $s_T$ and $d_L$ (pure quadratic dispersion) was
discussed by Eagles (1998b), but with the difference from the case
assumed here that a discrete lattice was considered in the
longitudinal as well as transverse directions. However, we have
recently found, based on work of Alekseev (2002), that for some of the
arrays considered there it is difficult to unambiguously define a Bose
condensation temperature because of the gradual way in which occupation
numbers of the ground state change with temperature.  This problem also
arises for the smallest arrays for a linear dispersion relation
(Alekseev 2001), but for this case the width of the transition is
proportional to $1/N_B^{\frac{1}{2}}$, where $N_B$ is the total number
of bosons, and is less than about 16\% of the condensation temperature
when $N_B>150$.

Condensation into the lowest-energy state cannot take place if the sum
of the boson occupation factors over all states except the lowest for a
chemical potential situated at the energy of the lowest state is
greater than the total number of bosons, $N_B$.  Below the temperature
at which this sum is equal to $N_B$, the occupation of the ground state
and of states in its neighbourhood will start to become macroscopic,
i.e. of the order of $N_B$, and so the equality of the sum with $N_B$
determines the condensation temperature $T_B$ in this weak sense (Rojas
1997).  For an average linear concentration $c$ of bosons per filament,
this criterion can be written as
\begin{equation}
\sum_{K_1,K_2,K_3}\frac{1}
{{\rm exp}[E_B(K_1,K_2,K_3)/k_BT_B]-1}
= cL n_T^2,
\end{equation}
where  the sum over $K_1$ and $K_2$ are sums from $i$, $j$ = 1 to $n_T$ of
$K_1=i\pi/D$ and $K_2=j\pi/D$, and the sum over $K_3$ is a sum from $k=1$
to infinity of $K_3=k\pi/L$, but the ground level $i=j=k=1$ is excluded.

As mentioned above, the ground-state occupation for small arrays
changes gradually, but we shall only present results for total numbers
of bosons equal to 150 or more.

\begin{center}
{\S 4. A CHOICE OF PARAMETERS WHICH CAN FIT THE DATA}
\end{center}

For superpolarons in elastomers with rotatable polar groups, the
potential felt by  electrons in the superpolaron arises from about
three oriented dipoles per electron surrrounding the superpolaron
string (Grigorov 1990). Even if arranged regularly, the
periodic part of the potential due to these surrounding dipoles
is expected to be small, and so the use of a continuum model in the
direction of the filaments is probably justified.

\vspace{0.5cm}
{\em $E(K)$ curve at $T=0$}

In one dimension at $T=0$ the coefficient of the linear term in the Cooper pair
energy at a center-of-mass wave vector $K$ is  $\hbar^2k_F/m_s$ (Casas
{\em et al.} 1998), where $m_s$ is the single-particle mass, while in
three dimensions it is half of this.  The case of an anisotropic
three-dimensional system has not been studied yet, and so we have to
make some assumptions.  For the direction along the channels where the
mass is smallest, we suppose that the one-dimensional result holds.
Hence, if the pair binding energy is large compared with the
Fermi energy, when the coefficient of the quadratic term is
approximately $\hbar^2/4m_s$ (Adhikari {\em et al.} 2000), then the
ratio $r$ of the coefficients of the linear to quadratic terms is
\begin{equation} 
r=4k_{FL},  
\end{equation}
where we have added the suffix $L$ to denote the longitudinal direction.
Using the relation
\begin{equation}
k_{FL}=\frac{1}{2}\pi c_F
\end{equation}
for one dimensional systems, where $c_F$ is the linear density of fermions,
and putting $c_F=2c$, with $c$ the linear concentration of bosons, 
we find that 
\begin{equation}
d_L=4\pi(ca_T).
\end{equation}

A proper calculation of the parameter $s_T$ in Eq.(1) would be
difficult.  In the Appendix we describe two suggestions which may give
approximate upper and lower limits for $s_T$ in terms of the ratio 
$W_T/(A_LK_{MT}^2)$, where 
$K_{MT}$ is the maximum tranverse
wave vector in the first Brillouin zone, i.e.
\begin{equation}
K_{MT}=\pi/a_T.
\end{equation}
In order to reduce the number of adjustable parameters by one we take
$s_T$ for a given $W_T$ and $p$ equal to the geometric mean of the
possible upper and lower limits.  
It turns out that this mean is independent of
$W_T$ and $p$ and is given by
\begin{equation}
s_T=2^{\frac{1}{2}},
\end{equation}
implying
\begin{equation}
p=W_T/[(1+2^{\frac{1}{2}})A_LK_{MT}^2].
\end{equation}

\vspace{0.5cm}
{\em $E(K)$ curve at finite $T$}

As far as we know, no $E(K)$ curves for Cooper pairs at finite
temperature have been published.  We expect that, when $T$ is greater
than the degeneracy temperature, i.e. when $k_BT_B>E_F$, 
with $E_F$ the Fermi energy, then the linear
term in the dispersion will no longer exist.  We make the simplest
assumption consistent with this, i.e. that the coefficient of both the
longitudinal and transverse linear terms decrease by a factor $F$
given by
\begin{equation}
F = 1-k_BT/E_F
\end{equation}
if $k_BT<E_F$, and $F=0$ when $k_BT>E_F$.

We still have four parameters to vary, and the only pieces of data we
have to fit are the ratio of the largest cross sections to give normal
conductivity at room temperature for channels of lengths 0.3 and 50
$\mu$m, and some results on limits of transition temperatures.  Thus we
cannot determine all four parameters.  We choose $a_T=2$ nm,
$c=0.5\times 10^7$ cm$^{-1}$, and $p=0.12$.  We then determine a value
of $M_L$ which can fit the fact that the ratio of the cross sections of
the channels with largest normal resistance at room temperature is
about four.  For the value of $c$ above, Eq.(7) gives
$k_{FL}=0.5\pi \times 10^7$ cm$^{-1}$.  Solution of Eq. (5) for
these parameters for channel lengths of 0.3$\mu$m and 50$\mu$m are
shown as a function of $n_T^2$ for $M_L=3.3 m_e$ in figure 2.
\begin{figure}
\rotatebox{270}
{\centerline{\includegraphics[width=12cm]{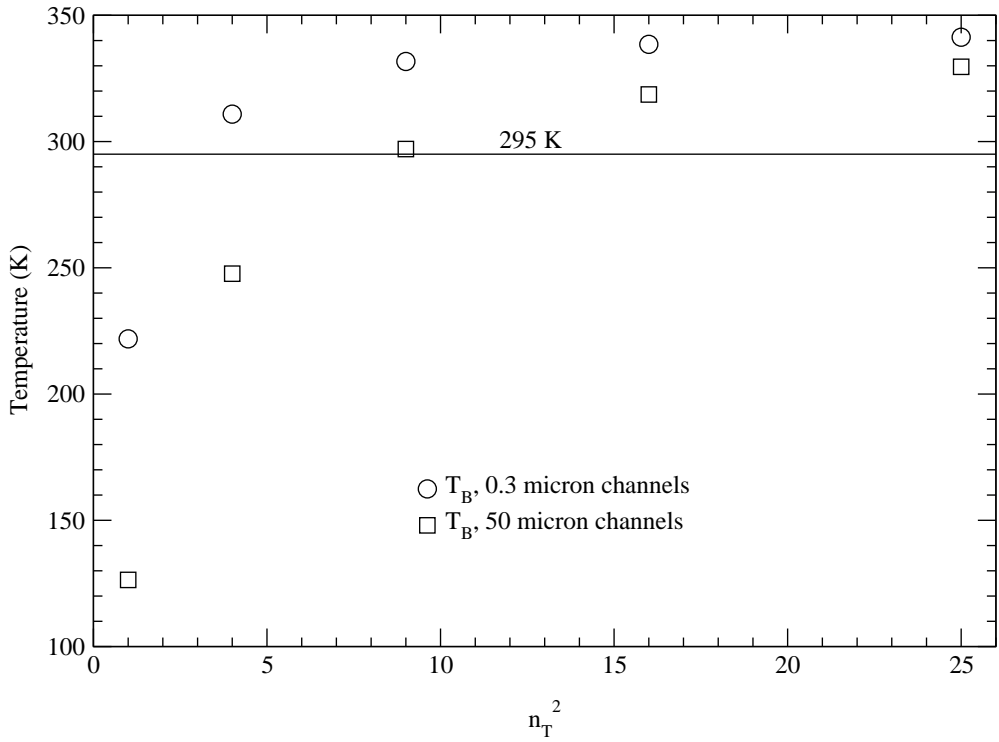}}}
\caption
{Condensation temperatures $T_B$ for two values of $L$ as a
function of $n_T^2$, with the assumptions $a_T=2$ nm,
$s_T=2^{\frac{1}{2}}$ at $T=0$, $M_L=3.3m_e$, $c=0.5\times
10^7$cm$^{-1}$, and $p=0.12$. At room temperature channels with
$n_T^2\leq 2$ are normal for $L=0.3\mu$m and channels with $n_T^2\leq
8$ are normal for $L=50\mu$m.}
\end{figure}
Interpolating between values of $n_T^2$ of 1, 4 and 9 for which
calculations have been done, we find that the 0.3$\mu$m channels are
normal at room temperature for $n_T^2$ up to 2, and the 50$\mu$m
channels are normal for $n_T^2$ up to 8, and so the ratio of the cross
sections of the lowest resistance normal channels for the thicker and
thinner films is four, in between the values of about five from figure 1
and of about two from Enikolopyan et al. (1989), as discussed in \S 2.

From figure 1 the lowest resistance of normal channels for the 0.3$\mu$m
film is about $2\times 10^4\Omega$, which according to figure 2
corresponds to two nanofilaments in the channel.  With $a_T=2$ nm, and
a channel length of $0.3\mu$m, the resistivity of the channel is
$0.53\times 10^{-4}\Omega$ cm.  With a boson density of
$ca_T^{-2}=1.25\times 10^{20}$ cm$^{-3}$, we deduce boson mobilities of
about  470 cm$^2$/Vs.  With a charge of $2e$ and a mass $M_L=3.3 m_e$,
this corresponds to a relaxation time $\tau=4.4\times 10^{-13}\rm{s}$
assuming diffusive transport.  For a single-particle mass of 1.65 $m_e$
and $k_F=0.5\pi\times 10^7$ s$^{-1}$, the velocity due to the linear
term in the dispersion is $1.1\times 10^7$cm s$^{-1}$.  For a pure
quadratic dispersion and a mass of 3.3 $m_e$, the thermal velocity at
room temperature $(k_BT/M_L)^{\frac{1}{2}}$ would be $0.37\times
10^{7}$ s$^{-1}$, and so the total velocity might be expected to be a
bit less than $1.5\times 10^7$ cm s$^{-1}$, corresponding to a
mean-free path $l\approx 70$ nm. The circumference of a channel with
two nanofilaments is 12 nm, so we may be on the ballistic side of the
quasi-ballistic diffusive transition according to statements of
Sch\"{o}nenberger {\em et al.}(1997).  If that is the case, we may
deduce that the conduction in the channel may be quasiballistic.  In
that case the resistance $R$ can be expected to be given by
$R=(R_Q/N_p)(L/l)$, where $R_Q=12.9$ k$\Omega$ is the quantum of
resistance, $N_p$ is the number of paths through the channel (we use
paths here instead of the usual channels as we have already used
channel in a different sense), $l$ is the mean-free path, and $L$ is
the channel length.  Taking $N_p=2$, this expression implies $l=100$ nm
for $R=2\times 10^4\Omega$ and $L=0.3\mu$m, not much different from the
value given by the diffusive transport calculation.

If there is any periodic part of the potential due to dipoles surrounding
the superpolaron, it is probably fairly weak, and so the bare mass of
the electrons in the superpolaron string is expected to be close to the
free-electron mass.  However, we expect that the pairing of electrons
in the superpolaron occurs by mediation of very high energy phonons
(Eagles 1994a) or plasmons (Eagles 1994b), and the binding energy of
pairs needs to be at least about 0.1 eV for pairs not to break up at
room temperature, or larger if the break up does not occur until higher
temperatures, as the higher $T_c$'s inferred from measurements with
{\em dc} (Enikolopyan {\em et al.} 1989) or pulsed-current measurements
(Demicheva {\em et al.} 1990) imply.  Thus the coupling with the
high-frequency modes may be moderately strong, and so the
single-particle masses and pair masses can be appreciably larger than
$m_e$ and $2m_e$ respectively.  The radius of a polaron due to
interaction with high-frequency modes of frequency $\omega$ will be of
the order of $(\hbar /2m_b \omega)^{\frac{1}{2}}$, where $m_b$ is the
bare electron mass. For $m_b\approx m_e$ this radius is about 0.3 nm
for $\hbar \omega \approx$ 0.37-0.39 eV intramolecular phonons
(McDonald and Ward 1961) or smaller for the larger $\hbar \omega$
expected for plasmons (Eagles 1994b),  but we expect that the two
polarons of a bipolaron will be separated by a larger amount
because the Coulomb repulsion between the two polarons will keep them
from coming too close.   Hence the pair mass $m_p$ may be close to
twice the single-particle mass, as implied by the use of the
Cooper-pair expression for the dispersion.

Note that the radius of a single polaron within a string and the 
pairing within a string due to interaction via high-frequency modes
arise from different processes from those which cause the formation of
the string itself.  This is analogous to theories of stripe formation
in high-$T_c$ cuprates, where the interactions which cause the stripes
to form are not generally thought to be the same as those which can
cause pairing of carriers within a stripe.  It is also worth noting
that the concept of superpolaron strings in polymers was introduced
many years before the concept of stripes in cuprates was considered.

Use of Eq.(7) implies that both spin states of the electrons are
equally populated.  Grigorov {\em et al.} (1990) have argued that the
superpolaron strings will be ferromagnetic.  However, we assume here
that the larger pairing energies for singlet pairing will overcome the
tendency of the superpolaron to be unstable when exchange energy is not
minimised by a ferromagnetic arrangement.

For $m_s=1.65 m_e$ and $k_F=0.5\pi\times 10^7$ cm$^{-1}$, the Fermi
energy $E_F=0.057$ eV, and so the reduction factor $F$ of the linear
terms in the Cooper pair dispersion at room temperature is 0.56.

\begin{center}
{\S 5. A POSSIBLE INCREASE OF CONDENSATION TEMPERATURE WHEN CURRENT FLOWS}
\end{center}

Figure 2 and calculations for larger $n_T$ imply that the limiting value
of $T_B$ (for large $n_T$) is not greater than about 350K
for the thicker films of figure 1.  Some films called Type 1 (Enikolopyan
{\em et al.} 1989) for which oxidation occurred slowly over several
years without UV irradiation did have $T_c$ close to room temperature.
Other films, Type 2, for which UV irradiation was used for oxidation,
had $T_c>429$ K when 0.5 A currents were passed through a channel
(Enikolopyan {\em  et al.} 1989), and probably greater than 700 K when
large pulsed currents were applied (Demicheva {\em et al.} 1990).  The
thicker films of figure 1 were of Type 2, and so might be expected to
have $T_c$ appreciably larger than room temperature when currents are
passed through channels.  We discuss a possible mechanism by which
$T_c$ can increase when current is applied.  The proposed mechanism
depends on the attractive force between filaments due to
current-current interaction when currents are passing through.
Because of the softness of the material, such forces can be expected to
decrease the distance between filaments, and hence decrease the
transverse masses of the channels.  This gives rise to an increase in
$T_B$.  However, when we put in figures for a channel of width of the
order of 1 $\mu$m, the rise in $T_B$ is much too low to explain the data.
Thus, in order to keep superconductivity at a given temperature when a
current is applied, we postulate that the current concentrates in a
smaller-width subchannel for which the current raises $T_B$ above the
relevant temperature.  This will be energetically favourable because
dissipation is avoided.  We show below that, when $a_T$ = 2 nm, $M_L =
3.3 m_e$, $(M_L/M_T)=0.12$, and $c=0.5\times 10^7$cm$^{-1}$, concentration of
a 0.5 A current in a subchannel of  initial lateral dimensions of about
0.24 $\mu$m is probably sufficient to raise $T_B$ of the subchannel to
above 429 K.

For an applied current of $I$ in a square subchannel with $n_T^2$
filaments, the current per filament, $I_f$, is $I/n_T^2$.  The pressure
on any plane of filaments not at the surface of the array does not
arise just from an external pressure at the surface, but is brought about by
the forces on all planes in the array further out from the center than
the plane being considered.  Then, for odd $n_T$, for a plane $m$
lattice constants from the center of the array, neglecting end effects,
the pressure corresponds to an effective force per unit length on a
filament at the center of the plane $F_{eff}(m)$ given by
\begin{equation}
F_{eff}(m)= (2I^2/a_Tn_T^4)
\sum_{i=m}^{q}\sum_{j=-q}^{q}\sum_{l=-q}^{q}
[(j-i)/[(j-i)^2+l^2)],
\end{equation}
where
\begin{equation}
q=0.5(n_T-1),
\end{equation}
and the triple sum excludes any terms with both $(j-i)=0$ and $l=0$.
Then the average pressure over the whole array $P_{av}$ is given by the
average of $F_{eff}(m)/a_T$ i.e. by 
\begin{equation}
P_{av}=[1/(q+0.5)](1/a_T)[0.5F_{eff}(0)+\sum_{m=1}^{q}F_{eff}(m)].
\end{equation}

Using this result and taking a channel with  $a_T=2$ nm, and
$D=0.242\mu$m, corresponding to $n_T=121$, we find, for an applied
current of 0.5 A (=0.05 emu), that the average pressure on filaments at
the centers of planes of the array is $2.4\times 10^6$ dynes/cm$^2$.
For OAPP films discussed by Grigorov {\em et al.} (1996), a pressure of
0.032 MPa gives rise to a reduction in thickness of about 3\%, i.e.
the compressive elastic modulus is about 1 MPa = 10$^7$ dynes/cm$^2$.
At higher pressures there is also a slow plastic flow, and so
calculations assuming the above elastic modulus will give a lower limit
to the reduction in thickness caused.  Thus we can expect the average
stress due to current-current interactions to produce at least about a
24\% average change in the filament separation $a_T$.  A 24\% decrease
is equivalent to a decrease by a factor of exp(-0.27).  For a small
overlap of wave functions between channels,  this percentage decrease
in $a_T$ will translate into a larger percentage increase of transverse
bandwidth, by a factor of exp(f$\times$0.27), where perhaps $f$
lies between 3 and 5. If $f=4$, then the ratio $p=M_T/M_L$ increases
by a factor of about 2.9. From our computer program to
solve Eq.(5), keeping other parameters as before, using $L=20 \mu$m
[the thickness of the particular film used was not stated in
Enikolopyan {\em et al.} (1989), but the results are not very sensitive
to the thickness], decreasing $M_T$ so that $M_L/M_T$ increases by a
factor of 2.9 from 0.12 to 0.35 for fixed $M_L$ increases $T_B$ for the
largest value for $n_T$ for which we have made calculations ($n_T=70$)
to above 429 K, as determined by experiments with 0.5 A currents
passing through a channel.

To calculate a drift velocity for a given current for a reduced $a_T$,
the carrier density goes up by the same factor as the current density
increases.  Thus we can calculate the drift velocity for the original
values of $a_T$.  A current of 0.5 A passing though a subchannel of
cross section (0.242 $\mu$m)$^2$ corresponds to a current density within
the subchannel of $8.54 \times 10^8$ A cm$^{-2}$,  smaller than
the order of magnitude estimate for the critical current density of $5
\times 10^9$ A cm$^{-2}$ made from pulsed-current measurements by
Demicheva {\em et al.} (1990).  For a boson density of $ca_T^{-2}=1.25\times
10^{20}$cm$^{-3}$, the drift velocity corresponding to a current
density of $8.54\times 10^8$ A cm$^{-2}$ is 2.13$\times 10^7 $cm s$^{-1}$.

For a single-particle mass $m_s=1.65 m_e$ and a Fermi wave vector
$k_F=0.5\pi\times 10^7$cm$^{-1}$, the linear term in the boson
dispersion corresponds to a velocity of $\hbar k_F/m_s=1.10\times 10^7$
cm s$^{-1}$, and thus the drift velocity corresponding to the quadratic
term in the dispersion must be $1.03\times 10^7$cm s$^{-1}$. For
$M_L=3.3 m_e$, we deduce that the quadratic term in the energy is 0.10
eV, that the wave vector $K$ corresponding to the drift velocity is
$2.9\times 10^7$ cm$^{-1}$, and that the linear term in the energy is
0.21 eV.   Thus the total pair binding energy is larger than 0.31 eV.
The bipolaron masses we deduce are quite sensitive to the boson
concentration, e.g. if we assume a concentration twice as large as used
above, the longitudinal boson mass deduced is increased to  above $10
m_e$.  For a given mass, the condensation temperature increases faster
than linearly with carrier concentration, partly because of the
decreasing effect of the factor $F$ of Eq.(12) as the concentration
increases.  Hence the mass to give a given condensation temperature
at a fixed concentration increases faster than linearly with carrier
concentration.

Note that Zhao (2003) has inferred energy gaps 2$\Delta$ of between
0.109 and 0.21 eV from tunnelling and Raman data in various carbon
nanotubes, for which he infers from many types of data that
superconducting $T_c$'s are sometimes greater than 600 K.  Our inferred
lower limit for the pair binding energy in OAPP is somewhat larger
than thought to exist in carbon nanotubes.  However, the masses we find are
larger in OAPP, and so a larger binding energy and a further
penetration into the Bose gas regime may occur in OAPP.

We consider that we have given a plausibility argument for a
significant increase in superconducting $T_c$ when a 0.5 A current
passes.  The case of much larger pulsed currents (Demicheva
{\em et al.} 1990) is more complicated, and will not be discussed
in detail here.  However we make a few remarks about this case.  First
it was argued in Eagles (1994b) that the critical currents of about 60 A 
observed by Demicheva {\em et al.} (1990) may have been the combined
total currents of several channels.  Nevertheless, if e.g. we assume a
a 20 A current in a channel of cross section $(1.002 \mu$m$)^2$
($n_T=501$), using Eq.  (13) and arguments similar to those for the
case of 0.5 A current, we find that the stress is larger by a factor of
$40^2\times$0.058 = 93 than for the current concentrated in a
subchannel of diameter 0.242 $\mu$m.   For such large stresses, strong
non-linear effects in the stress-strain curve will occur.  We expect
that when the stress is large enough to bring atoms sufficiently close
to each other, then the elastic constants will increase to values
similar to those of more normal solids, i.e. by a factor of the order
of $10^6$.   Thus there will be a saturation of the strain, but it
would be difficult to make any quantitative estimates.

Two other complications in the pulsed current case are: (i) When
currents are close to those required to destroy superconductivity,
there will be a competition between the increase in $T_B$ due to
increase of the transverse bandwidth by compressive stresses, and a
lowering of $T_B$ due to  approach to the depairing current [cf.
Kunchur {\em et al.} (2003)]. (ii) Another complication is the question
of time scales for increases and decreases of $a_T$ which will play a
part in repetitive pulsed-current measurements.

\begin{center}
{\S 6. DISCUSSION}
\end{center}

There are several arbitrary parameters in our theory besides the
assumptions about the boson dispersion, and only three pieces of
quantitative information to fit, viz. the approximate ratio of the
lowest normal-state resistance of channels in films of two thicknesses,
plus minimum values of room temperature and 429 K for superconducting
transition temperatures at low currents and with 0.5 A currents
flowing.  Thus the best that can be said is that we have given a
possible explanation for the results.

The $E(K)$ curves we have used for bosons in the longitudinal direction
are based on results for Cooper pairs with an instantaneous
electron-electron attraction.  In that model the mass of the pair is
twice that of the single particle whatever the strength of the
electron-electron attraction.  In bipolaron theories on the other hand,
there are indications [see e.g. Macridin {\em et al.} (2004)] that the
pair mass is only approximately equal to that of the single particles
if the bipolaron binding energy is very small.  However, we think
that such a result could be modified when two different frequencies of
excitations mediating the electron-electron attraction are involved.
Then a small interaction mediated by a very high-frequency electronic
excitation can affect the bipolaron binding energy appreciably while
only giving a small change to the effective mass.

Our chosen $E(K)$ curves for the transverse direction and the
temperature dependence of these curves for both directions are
guesses.  Although our chosen value for the value of the coefficient of
the linear term for the transverse direction could be far from correct,
if we choose another value for this coefficient which is not too small,
then fairly similar results $T_B$ can be obtained by adjusting the
transverse mass to compensate.  As an example, if we change $s_T$ at
$T=0$ from $2^{\frac{1}{2}}$ to $0.5\times2^{\frac{1}{2}}$ and increase
$p$ from 0.12 to 0.155, $T_B$ for $n_T$ between two and ten for both
the shorter and longer channels is changed on average by well under
1\%.

Another point which we mention is the following.  The minimum number of
nanofilaments to support superconductivity at room temperature deduced
in \S 4 (nine for the longer channels) is less than the 30 deduced from
a model for the metamagnetism given in Eagles (1998a).  It is possible
to increase the numbers of filaments required for superconductivity at
room temperature in the model of the present paper by either increasing
$M_L$ or increasing $M_T$, but then a larger fractional decrease of
$M_T$ when 0.5 A currents are flowing is required to increase $T_B$ to
above 429 K.  As an example, if we leave $M_L$ as before but increase
$M_T$ to make $p=0.062$ at low currents, then the largest values of
$n_T^2$ for which the channels are normal at 295 K are $n_T^2=8$ for
0.3$\mu$m channels and $n_T^2=25$ for 50$\mu$m channels.  This gives a
ratio of cross sections of 3.1, not far from the  mean of values of
about five inferred from figure 1 and of about two from Enikolopyan
{\em et al.} (1989), and a value of $n_T^2$ larger than 25 is needed to
obtain superconductivity in the longer channels, in agreement with that
of 30 required in the model of Eagles (1998a).   If we now suppose that
the 0.5 A current concentrates in a subchannel of width 0.202$\mu$m,
corresponding to $n_T=101$, then the decrease of $a_T$ caused by the
current is calculated to be by a factor of 0.66 = exp(-0.42).  If we
assume that this corresponds to an increase of $p$ by a factor
exp(0.42$\gamma$), with $\gamma$ taken as 4.2, then this gives an
increase of $p$ from 0.06 to 0.35, sufficient to increase $T_B$ to
above 429 K for large $n_T$.  However the pair binding energy deduced
from the larger drift velocities caused by concentrating the currents
in a subchannel of smaller diameter than before now has to be larger
than 0.76 eV, assuming the $E(K)$ curve for pairs retains the same form
up to such rather large energies.  This need not necessarily be the
case.  For the Cooper-pair problem with a finite-range interaction in
two dimensions, it has been shown that, for large pair binding
energies, a secondary minimum in $E(K)$ can develop, with a larger
quadratic contribution to the energy for departures of $K$ from its
value at the minimum, (corresponding to a smaller mass at this
secondary minimum) compared with that at $K=0$ (Adhikari {\em et al.}
2000).

As an alternative to concentrating the current in a channel of
smaller-diameter than before, we could consider the possibility that
carrier concentrations in the channels may be increased by injection
from the contacts when a current is applied, but we do not have a
quantitative theory for this.  A decrease of separation in distances
between channels as current is increased was inferred from Josephson effect
measurements at low temperatures in films of
poly(phthalidylidenebiphenylene) (Ionov {\em et al.} 2002), but this
does not give information about carrier concentrations within the
channels.  Also Skaldin (1991) analysed results on channels through
thin films of polydiphenylenephthalide (Skaldin {\em et al.} 1990) to
infer that cross sections of channels increase in proportion to applied
currents, and the conductance of the channels increases in proportion
to the cross sections, which implies that there is no increase in
carrier concentration in the channels in this material.

The model of Eagles (1998a) involved quite a lot of assumptions,
including that of spontaneous high currents circulating  both within
linear channels and round channels which form closed loops.  This
second type of assumption was not required in a recent interpretation
(Eagles 2002) of large diamagnetism at low fields in films of oxidised
atactic polypropylene (Rogachev and Grigorov 2000), and it is possible
that the first type of assumption may not be necessary to explain the
metamagnetism.  Thus it would not be very surprising if the model of
Eagles (1998a) did not give correct results for minimum numbers of
nanofilaments required to support superconductivity.  However, if we
were to suppose that the model described there for metamagnetism is
qualitatively correct, then we would expect that spontaneous currents
would also flow round channels which form closed loops, and so a
different explanation from that given in Eagles (2002) would be
required to explain details of the magnetic susceptibility versus field
curves in two samples of films of OAPP analysed in that paper.

More experimental results of the type shown in figure 1 would be of great
interest, especially if they could be combined with measurements of the
diameters both of the channels of medium resistance  and of the
superconducting channels.  Many other types of experiments which might
be worthwhile to perform on films of oxidised atactic polypropylene and
other polymers are suggested in \S 3 of Eagles (1998a).

\S 7. MORE GENERAL DEPENDENCE OF $T_B$ ON PARAMETERS

It is hoped that a paper with coauthors giving a more general
discussion of how $T_B$ depends on various parameters will be written
later.  However, to give some ideas about this in the meantime, we have
done various calculations with all except one parameter fixed at values
given in the caption to figure 2, while one is varied, for several values
of $n_T$.  A typical value of channel length $L=20\mu$m is considered if $L$
is not the parameter to be varied.  We find:  
\newline 1. $T_B$ is proportional to the reciprocal of the longitudinal
mass.  However, bipolaron binding energies will decrease as masses
decrease, making it more difficult to get on the Bose gas side of the
BCS-Bose gas transition for small masses.  \newline
2. $T_B$ also increases as the transverse mass decreases but not so
fast.   $T_B$ is approximately proportional to $M_T^{-0.184}$ for values of
$M_T/M_L$ that lie between 0.04 and 0.3  for $L=20\mu$m, $n_T=10$ and other
parameters as in the caption to figure 2.  However, when $M_T$ becomes a
significant fraction of $M_L$, then the quasi one dimensionality is
lost, and so it may be less easy to obtain a large bipolaron binding
energy.   The increase in $T_B$ as $M_T$ decreases (subject to the
limitation mentioned above) implies that $T_B$ will increase rather
faster as the transverse lattice constant $a_T$ decreases, assuming
$M_T\propto a_T^{-\nu}$, with $\nu$ in the range of  3 to 5.\newline
3. $T_B$ increases faster than linearly as the boson concentration
increases.  For $n_T=10$, $L=20\mu$m and other parameters as in the
caption to figure 2, $T_B$ increases with $c$ for $ca_T$ between 0.04 and
0.5 approximately as a power between 1.61 and 1.75, with the
exponent rising slowly as $c$ increases.\newline   
4. We have performed calculations for $n_T$ from 1 to 10 for values of
$L$ from 0.1 to 1000 $\mu$m for other parameters as in the caption to
figure 2.  For $n_T=1$, $T_B$ decreases monotonically with increasing
$L$ and decreases by 62\% as $L$ increases by four orders of
magnitude.   For other values of $n_T$, $T_B$ goes through a 
maximum as $L$ begins to increase, and then decreases, with the rate of
decrease slower for larger $n_T$, e.g. as $L$ changes from 100 to 1000
$\mu$m, $T_B$ decreases by 4.0\% for $n_T=5$ and by 2.9\% for
$n_T=10$.   These slow decreases do not appear to be sufficient to
account for the fact that room-temperature superconductivity has never
been reported through films of thickness greater than 100$\mu$m
(Enikolopyan {\em et al.} 1989).  This is more likely to be associated with
the difficulty of forming stable long and straight conducting channels (Grigorov
1992).
\newline
5. As the zero temperature value of the ratio of linear to quadratic
contributions to the transverse bandwidth is varied from 0 to 3, for
$n_T=10$, $L=20\mu$m and other parameters as in the caption for figure 2,
$T_B$ increases by 24\%.\newline
6. Figure 2 gives examples of how $T_B$ depends on $n_T$.  Compared with
the case of pure quadratic dispersion discussed by Eagles (1998b), the
initial rate of rise of $T_B$ with $n_T$ is much faster for the
combined linear plus quadratic dispersion.

\S 8. WHAT IS NEEDED FOR HIGH $T_c$ BESIDES QUASI ONE DIMENSIONALITY?

We have suggested that electron-phonon interactions and/or
electron-plasmon interactions are the cause of the pairing of electrons
in channels through OAPP.  However, other polar quasi one-dimensional
materials can be expected to have electron-phonon interactions of
comparable strength, and, for appropriate current carrier
concentrations,  we might at a first guess think that electron-plasmon
interactions could comparable magnitude to those in channels through
OAPP.  So why have not many other types of quasi one-dimensional
materials shown superconductivity at very high temperatures?  We do not
have a definite answer to this question, but make a few tentative
suggestions here.

First we note that the only other two quasi one-dimensional materials
that we are aware of for which room-temperature superconductivity has
been claimed besides those discussed in this paper are, (i) carbon
nanotubes (Tsebro {\em et al.} 1999, Zhao and Wang 2001, Zhao 2003), and
(ii) powdered mixtures of PbCO$_3$.2PbO $+$ Ag$_2$O (Djurek {\em et al.}
2001).  The structure of the superconducting components of the system
studied by Djurek {\em et al.} has been suggested to contain well separated
Ag-O chains which are thought to be the main channels for possible
superconductivity in this system.  The only atoms in carbon nanotubes
are carbons, and the linear chains thought to give the
superconductivity in the system of Djurek {\em et al.} are composed of silver
and oxygen atoms.  In both of these systems the quasi one-dimensional
conduction takes place by electrons which do not have a $d$-electron
component.  Further, for OAPP and other elastomers, according the the
theory of Grigorov (1990,1991), nanofilaments from which the conducting
channels are composed pass through spaces in the polymeric system, and
so the electrons in the nanofilaments of the channels avoid all atoms,
and do not involve $d$-electrons or any other atomic orbitals.   Thus,
on the basis of what types of materials are thought by some to show
room-temperature superconductivity, we make our first hypothesis that,
besides quasi one-dimensional conduction, the conduction  electrons
must not have a $d$-electron component.  If $d$ electrons take part in
the conduction, Coulomb repulsion can be expected to be large, and so,
for given strengths of electron-phonon and electron-plasmon
interactions, it will be more difficult to obtain large pair binding
energies, assuming $s$-pairing symmetry.  (We do not discuss $d$-wave
superconductivity here, for which on-site Coulomb repulsion does not
cause a problem).

However, the above criterion does not tell the whole story, because
there are quasi one-dimensional polar organic materials which do not
involve $d$-electrons but which only show superconductivity at low
temperatures, or instead form charge-density waves.  The squares of
matrix elements of electron-plasmon interactions are proportional to
$\epsilon_h^{-3/2}$, where $\epsilon_h$ is the high-frequency dielectric
constant of the material (see e.g. Bose and Gayen, 2004) and the
squares of the matrix elements for Fr\"{o}hlich interactions
interactions with longitudinal optical phonons are proportional to
$(1/\epsilon_h-1/\epsilon_s)$ (Fr\"{o}hlich 1954), where $\epsilon_s$
is the static dielectric constant.  Thus a low high-frequency
dielectric constant is important for strong interactions mediated via
plasmons, and for interactions mediated by longitudinal optical phonons
a low $\epsilon_h$ and a fairly high $\epsilon_s$ are necessary.  The
high-frequency dielectric constant of polypropylene is  $1.57^2\approx 2.2$
(Brandup and Immergut 1989), and an effective dielectric constant for
graphite has been reported to be 1.4 (Egger and Gogolin 1998).  The value of
2.2 for polypropylene is not much smaller than the value of
$\epsilon_h$ for several organic conductors and superconductors
(J\'{e}rome and Schultz 1982).  Thus, other things being equal,
electron-plasmon interactions might be expected to be of similar
magnitude in OAPP and some quasi one-dimensional organic
superconductors such as the (TMTSF)$_2$X series.  The static dielectric
constant $\epsilon_s$ is quite high in films of OAPP having
superconducting channels.  The increase of $\epsilon_s$ during heat and
ultra-violet treatment is associated with an increase of rotatable
dipolar groups.  However, when the concentration of dipolar groups
becomes sufficiently large, more atoms become ionised (Grigorov 1990),
and so at least part of the increase in $\epsilon_s$ should help to
increase the polarisation associated with lattice vibrations, and hence
to increase interactions between electrons and longitudinal optical
phonons.  We also note that some of the intramolecular vibrations in
polypropylene have very high frequencies (McDonald and Ward 1961), and
so contributions to bipolaron binding energies from interactions with
these phonons will be relatively large for a given strength of
electron-phonon interation.  We have not found information on
$\epsilon_s$ in the (TMTSF)$_2$X family yet, but the charges on the
whole molecules involved are or of magnitude half or one electronic
charge, and so the average charge density over the large (TMTSF)
molecules is expected to be small. Thus the polarisation density from
their motion and hence $\epsilon_s$ may not be very
large.  Hence electron-electron interactions mediated by longitudinal
optical phonons may be significantly larger in OAPP than in
(TMTSF)$_2$X.  Another factor that may reduce pairing energies is the
probable non-negligible on-site Coulomb repulsion for
$p$-electrons in the (TMTSF)$_2$X materials, and the possible
interference of repulsive interactions via magnetic excitations  with
attraction via phonons and plasmons.   Although OAPP has some unusual
magnetic properties besides fairly large diamagnetism at low fields
(Smirnova et al., 1998, Enikolopyan {\em et al.} 1989, Grigorov
{\em et al.} 1996, Rogachev and Grigorov 2000), we are not aware of any
magnetic excitations in this material which might mediate an
electron-electron repulsion.  Another factor to take into account is
competition between superconducting pairing and instabilities
associated with charge- or spin-density waves, which are more likely to
be of importance for electrons in a tight-binding type of band
structure rather than for the nearly-free bare electrons which we think
occur in the conducting channels through films of OAPP.  A final
property required for high condensation temperatures is a fairly low
bare mass, so that bipolaron masses need not necessarily be excessively
large.

So, based on arguments given in the last two paragraphs, we tentatively
suggest eight properties which are helpful for very high-temperature
$s$-wave superconductivity.  These are:\newline
1. Quasi one-dimensional conducting channels.  The restricted
dimensions help bipolaron formation.  Mourachkine (2004) is also of the
opinion that quasi-one dimensionality is a requirement for
room-temperature superconductivity, but suggests that bisolitons are
probably the type of pairs involved.  This may well be the case in some
materials. \newline 
2. In order to keep Coulomb repulsion low, conduction electrons in
these channels should not have any $d$-orbital component, and should
preferably be electrons which avoid atoms altogether as in channels in
OAPP.\newline
3. The quasi one-dimensional channels should lie in a background
material with a low $\epsilon_h$ in order to help to give large
electron-plasmon interactions.\newline
4. The background material should also have a high $\epsilon_s$, so
that interactions with longitudinal optical phonons can be large.  The
extent to which the increase in $\epsilon_s$ during heat and
ultra-violet treatment of atactic polypropylene causes increases
in electron-phonon interactions is not known at present. This is because
the increase in $\epsilon_s$ is associated with increase of rotatable
dipolar groups, but also with increases in numbers of ionised atoms.
\newline
5. The presence of very high-frequency phonons is an advantage, as this
permits larger bipolaron binding energies for a given strength of
electron-phonon interaction.\newline
6. There should be a lack of competition with other forms of order
than superconductivity, such as charge- or spin-density waves.  Such
competition is less likely to occur if the bare electrons in the
channels are closer to being free-electron like than to being
describable by tight-binding types of band structure.  \newline
7. There should be a lack of competition between repulsive interactions
mediated via magnetic excitations and attractive interactions mediated
by plasmons and phonons.\newline
8. A fairly low bare-electron mass in the channels is required in
order to give a chance for the Bose-Einstein condensation temperature
of the pairs to be large.

OAPP has at least seven of these properties, and possibly all eight,
although number 7 above is uncertain.  Carbon nanotubes do not satisfy
property 4, but this is probably compensated by the unusually low
effective $\epsilon_h$, making electron-plasmon interactions unusually
strong.  We do not have enough information about the compound studied
by Djurek et al. (2001) to know how many of the above eight properties
are satisfied for this compound apart from the first two.

\begin{center}
{\S 9. CONCLUSIONS}
\end{center}

We have given a plausible explanation for the fact that the minimum
resistance of normal channels through oxidised atactic polyproplene
(OAPP) films increases with film thickness more slowly than linearly.
The explanation is based on calculations of the condensation
temperature $T_B$ of bosons in arrays of filaments when the bosons have
a mixed linear plus quadratic dispersion, as indicated by published
results for the Cooper-pair problem.  The condensation-temperature
calculations show that the minimum cross section of arrays which can
support superconductivity at a given temperature increases as the
length of the array increases.  To explain the observations in detail
with the model, we need to assume that, even for large tranverse
dimensions of the filament arrays, the superconducting $T_c$ at low
currents is not more than about 350 K.  To explain higher $T_c$'s
inferred from data involving higher applied currents, we need to
postulate that $T_c$ increases as the current increases, at least for
not too large currents.  A possible reason for this is shown to be
related to the fact that current-current interactions produce
compressive stresses which cause significant compression of the array
in the elastomer, atactic polypropylene, and that this gives rise to
larger transverse bandwidths with an associated increase in $T_B$.
However, in order to explain the results quantitatively, we need to
suppose that the applied current concentrates in a subchannel of
lateral dimensions of about $0.24\mu$m, smaller than the probable
diameters of the whole channels, of the order of 1 $\mu$m (Demicheva
{\em et al.} 1990).

\begin{center}
{APPENDIX}
\end{center}

It would be difficult to do a proper calculation of the parameter $s_T$
of equation (1).  In this Appendix we present two types of calculations which
probably give upper and lower limits for $s_T$.

For a simple anisotropic system with no cut offs in wave vectors, we
guess that the ratio $r$ of coefficients of linear and quadratic
terms for transverse motion would be given by 
\begin{eqnarray}
r=2k_{FT}=2k_{FL}(m_T/m_L)^{\frac{1}{2}},\;\;\;\;\;\;\;\;(A1)\nonumber
\end{eqnarray}
where $m_L$ and $m_T$ are the longitudinal and transverse masses for
single particles.  The factor of two in equation (A1) instead of four given
in equation (6) arises because of the three dimensionality of the system.
Equation (A1) implies
\begin{eqnarray}
s_T=2p^{-\frac{1}{2}}(k_{FL}/k_{MT}),\;\;\;\;\;\;\;\;\;\;\;\;\;\;\;{\rm (A2)}
\nonumber
\end{eqnarray}
provided $m_L/m_T=M_L/M_T=p$.  Here $k_{MT}=\pi/a_T$.  

However, when $k_{MT}$
is small compared with $k_{FT}$ given above, then, for most
values of the longitudinal component of the single-fermion wave vector,
the transverse wave vector is limited by the zone edge value
$k_{MT}$.   The fraction $f$ of longitudinal wave vectors for which the
transverse wave vector at the Fermi surface is less than $k_{MT}$
is given by
\begin{eqnarray}
f=1-(1-E_{MT}/E_F)^{\frac{1}{2}},\;\;\;\;\;\;\;\;\;\;\;\;{\rm (A3)}\nonumber
\end{eqnarray}
where $E_{MT}=(\hbar^2/2m_T)k_{MT}^2$.  If $E_{MT}/E_F$ is fairly small,
then we can approximate $f$ as 
\begin{eqnarray}
f \approx (1/2)(E_{MT}/E_F)=(p/2)(k_{MT}/k_{FL})^2,\;\;\;\;\;{\rm (A4)}\nonumber
\end{eqnarray}
if $m_L/m_T= M_L/M_T=p$.   
For those longitudinal wave vectors for which the transverse wave
vector is limited by the zone edge, we do not expect any contribution
to the linear term in the Cooper pair dispersion provided that any gap
at the zone edge is large compared with the pair binding energy.  In
that case we infer from equations (A2) and (A4) that $s_T$,
the ratio of linear and quadratic contributions to the transverse pair
bandwidth is given by
\begin{eqnarray}
s_T\approx p^{\frac{1}{2}}(k_{MT}/k_{FL})\;\;\;\;\;\;\;\;\;\;\;\;\;\;\;\;\;{\rm (A5)}
\nonumber
\end{eqnarray}

The assumption that the excitations from the parts of the Brillouin
zone where the transverse component of the Fermi wave vector extends
to the zone edge do not contribute to the linear terms in the Cooper
pair dispersion depends on any gap at the zone edge being large
compared with the pair binding energy.  In the opposite limit when the
pair binding energy is large compared with such gaps, we expect
that such gaps do not have much effect.  So, in that case $s_T$ might
be given by equation (A2).  Thus equation (A2) probably gives an upper limit for
$s_T$ and equation (A5) a lower limit.  Since we do not which of the two
limits is a better approximation, we use the geometric mean of the two,
which turns out to have the simple value
\begin{eqnarray}
s_T=2^{\frac{1}{2}}.\;\;\;\;\;\;\;\;\;\;\;\;\;\;\;\;\;\;\;\;\;\;\;\;\;{\rm
(A6)}\nonumber
\end{eqnarray}
This is the result we assume for $T=0$.  Our numerical calculations
indicate that, if we use different values of $s_T$ which are not very
small, then changes in $s_T$ can be approximately compensated for by
appropriate changes in $p$.

\begin{center}
{REFERENCES}
\end{center}
\noindent
ADHIKARI, S.K., CASAS, M., PUENTE, A., RIGO, A.,  FORTES, M.,
SOL\'{I}S, M.A.,  DE LLANO, M.,  VALLADARES, A.A., and  ROJO, O.,
2000, {\em Phys. Rev. B} {\bf 62},  8671.\newline
ALEKSEEV, V.A., 2001, {\em Zh. \'{E}ksp. Teor. Fiz.} {\bf 119}, 700 [{\em JETP}
{\bf 92}, 608].\newline
ALEKSEEV, V.A., 2002, {\em Zh. \'{E}ksp. Teor. Fiz.} {\bf 121}, 1273
[{\em JETP} {\bf 94}, 1091].\newline
ARKHANGORODSKI\u{I}, V.M., GUK, E.G., EL'YASHEVICH, A.M., IONOV, A.N.,
TUCHKEVICH, V.M., and  SHLIMAK, I.S., 1989, {\em Dokl. Acad. Nauk,
SSSR} {\bf 309}, 603  [{\em Sov. Phys. Doklady} {\bf 34},
1016].\newline
ARKHANGORODSKI\u{I}, V.M., IONOV, A.N.,  TUCHKEVICH, V.M.,
and  SHLIMAK, I.S., 1990, {\em Pis'ma Zh. Eksp. Teor. Fiz.} {\bf 51}, 56
[{\em JETP Lett.} {\bf 51}, 67].\newline 
BOSE, S.M., and GAYEN, S., 2004, {\em cond-mat}/0402531. {\em Phase
Transitions}, to be published. \newline
BRANDRUP, J., and IMMERGUT, E.H., 1989, in {\em Polymer Handbook}
(New York, Wiley).\newline
CASAS, M., RIGO, A., DE LLANO, M., ROJO, O., and SOL\'{I}S., M.A., 1998, {\em 
Phys. Lett. A} {\bf 245}, 55.\newline
DEMICHEVA, O.V., ROGACHEV, D.N., SMIRNOVA, S.G., SHKLYAROVA, E.I.,
YABLOKOV, M.Yu., ANDREEV, V.M., and GRIGOROV, L.N., 1990, {\em Pis'ma
Zh. Eksp. Teor.  Fiz.} {\bf 51}, 228 [{\em JETP Lett.} {\bf 51} 258].\newline
DJUREK, D., MEDUNI\'{C}, Z., TONEJC, A.,  and PALJEVI\'{C}, M., 2001,
{\em Physica C} {\bf 351}, 78.\newline
EAGLES, D.M., 1969, {\em Phys. Rev.} {\bf 186}, 456.\newline
EAGLES, D.M., 1994a, {\em J. Supercond.} {\bf 7}, 679.\newline
EAGLES, D.M., 1994b, {\em Physica C} {\bf 225}, 222; erratum
1997, {\em Physica C} {\bf 280}, 335.\newline
EAGLES, D.M., 1998a, {\em J. Supercond.} {\bf 11}, 189.
See footnotes in Eagles (2002) for information about some errors in this 
paper. \newline
EAGLES, D.M., 1998b, {\em Physica C} {\bf 301}, 165.\newline
EAGLES, D.M., 2002, {\em J. Supercond.} {\bf 15}, 243.\newline  
EGGER, R., and GOGOLIN, A.O., 1998, {\em Eur. Phys. J. B} {\bf 3}, 281.\newline
ENIKOLOPYAN, N.S., GRIGOROV, L.N., and SMIRNOVA, S.G.,
1989, {\em Pis'ma Zh. Eksp. Teor. Fiz.} {\bf 49},  326 [{\em JETP Lett.}
{\bf 49} 371].\newline 
FR\"{O}HLICH, H., 1954, {Adv. Phys.} {\bf 3}, 354.\newline
GRIGOROV, L.N., 1990, {\em Makromol. Chem., Macromol. Symp.}  
{\bf 37}, 159.\newline 
GRIGOROV, L.N., 1991, {\em Pis'ma Zh. Tekh. Phys.} {\bf 17} (9-10), 45
[{\em Sov. Tech. Phys. Lett.} {\bf 17}, 368].\newline
GRIGOROV, L.N., 1992, {\em Vysokomol. Soedin A} {\bf 34} (9), 74 [{\em
Polymer Science} {\bf 34}, 772].\newline
GRIGOROV, L.N., 1998, {\em Phil. Mag. B} {\bf 78}, 353.\newline
GRIGOROV, L.N., and SMIRNOVA, S.G., 1988, Deposited Article No. 2381-V88,
All-Union Institute for Scientific and Technical Information, 23 March.\newline
GRIGOROV, L.N., ANDREJEV, V.M., and SMIRNOVA, S.G., 1990, {\em Makromol.
Chem., Macromol. Symp.} {\bf 37}, 177.\newline
GRIGOROV, L.N., DEMICHEVA, O.V., and SMIRNOVA, S.G., 1991,
{\em Sverkhprovodimost' (KIAE)} {\bf 4}, 399 [{\em Superconductivity, Phys.
Chem. Tech.} {\bf 4}, 345].\newline
GRIGOROV, L.N., ROGACHEV, D.N., and KRAEV, A.V., 1993, {\em Vysokomol.
Soedin.} {\bf 35}, 1921 [{\em Polymer Science} {\bf 35}, 1625].\newline
GRIGOROV, L.N., DOROFEEVA, T.V., KRAEV, A.V., ROGACHEV, D.N., 
DEMICHEVA, O.V.,  and SHKLYAROVA, E.I., 1996, {\em Vysokomol. Soedin. A} 
{\bf 38}, 2011. [{\em Polymer Science A} {\bf 38}, 1328].\newline
IONOV, A.N., LACHINOV, A.N., and RENTZSCH, R., 2002, {\em Pis'ma Zh.
Tekh. Fiz.} {\bf 28} (14), 69 [{\em Tech. Phys. Lett.} {\bf 28}, 608].\newline
J\'{E}ROME, D., and SCHULTZ, H.J., 1982, {\em Adv. Phys.} {\bf 31}, 299.
\newline
KUNCHUR, M.N., LEE, S-I., and KANG, W.N., 2003, {\em Phys. Rev. B} {\bf
68}, 064516 \newline
LANGER, J.S., and AMBEGAOKAR, V., 1967, {\em Phys. Rev.} {\bf 164}, 498.
\newline
MACRIDIN, A., SAWATZKY, G.A., and JARRELL, M., 2004, {\em Phys. Rev.} 
{\bf 69}, 245111.
\newline
McCUMBER, D.E., and HALPERIN, B.I., 1970, {\em Phys. Rev. B} {\bf 1}, 1054.
\newline
McDONALD, M.P., and WARD, I.M., 1961, {\em Polymer} {\bf 2}, 241.\newline
MOURACHKINE, A., 2004, {\em Room-Temperature Superconductivity}
(Cambridge, Cambridge International Science Publishing).\newline
ROGACHEV, D.N., and GRIGOROV L.N., 2000, {\em J. Supercond.} {\bf 13}, 947.
\newline
ROJAS, H.P., 1997, {\em Phys. Lett. A} {\bf 234}, 13.\newline
SCH\"{O}NENBERGER, C., BACHTOLD., A., STRUNK, C., SALVETAT, J.-P., and
FORR\'{O}, L., 1999,
{\em Appl. Phys. A} {\bf 69}, 283.\newline
SKALDIN, O.A., 1991, {\em Pis'ma Zh. Tekh. Fiz.} {\bf 17} (19-20),
64 [{\em Sov. Tech. Phys. Lett.} {\bf 17}, 705].\newline
SKALDIN, O.A., ZHEREBOV, A. YU., LACHINOV, A.N., CHUVYROV, A.N.,
and DELEV, V.A., 1990, {\em Pis'ma Zh. Eksp. Teor. Fiz.} {\bf 51} 141
[{\em JETP Lett.} {\bf 51}, 159].\newline
SMIRNOVA, S.G., DEMICHEVA, O.V., and GRIGOROV, L.N., 1988, {\em Pis'ma
Zh. Eksp. Teor. Fiz.} {\bf 48}, 212 [{\em JETP Lett.} {\bf 48}, 231].
\newline
TSEBRO, V.I., OMEL'YANOVSKI\u{I}, O.E., and MORAVSKI\u{I}, A.P., 1999, 
{\em Pis'ma Zh. Eksp. Teor. Fiz.} {\bf 70}, 457 
[{\em JETP Lett.} {\bf 70}, 462].\newline
ZAIKIN, A.D., GOLUBEV, D.S., VAN OTTERLO, A., and ZIM\'{A}NYI, G.T., 1997,
{\em Phys. Rev. Lett.} {\bf 78}, 1552.\newline
ZHAO, G.M., and WANG, Y.S., 2001, {\em cond-mat}/0111268.\newline  
ZHAO, G.M., 2003, {\em cond-mat}/0307770.

\end{document}